\documentclass[twocolumn,showpacs,preprintnumbers,amsmath,amssymb,pra,superscriptaddress]{revtex4}
\usepackage{graphicx}
\usepackage{epstopdf}
\usepackage{dcolumn}
\usepackage{bm}
\usepackage{amsmath}
\usepackage{amssymb}
\usepackage{mathrsfs}
\usepackage{amsfonts}
\usepackage{color}

\begin{document}

\title{Metastability in the driven-dissipative Rabi model}
\author{Alexandre Le Boit\'e, Myung-Joong Hwang, and Martin B. Plenio}
\affiliation{Insitut f\"ur Theoretische Physik and IQST, Albert-Einstein-Allee 11, Universit\"at Ulm, D-89069 Ulm, Germany}

\begin{abstract}
We explore the long-time dynamics of Rabi model in a driven-dissipative setting and show that, as the atom-cavity coupling strength becomes larger than the cavity frequency, a new time scale emerges. This time scale, much larger than the natural relaxation time of the atom and the cavity, leads to long-lived metastable states susceptible to being observed experimentally. By applying a Floquet-Liouville approach to the time-dependent master equation, we systematically investigate the set of possible metastable states. We find that the properties of the metastable states can differ drastically from those of the steady state and relate these properties to the energy spectrum of the Rabi Hamiltonian.
\end{abstract}

\maketitle
\section{Introduction}
In the context of cavity quantum electrodynamics (QED), a common way to probe the quantum nature of the interaction between light and matter is to drive the system with a classical light field and record the statistics of the photons emitted from the cavity. For example, a sub-Poissonian statistics of output photons is an important evidence of effective photon-photon interactions induced by the atom-cavity coupling~\cite{Walls:2007}. Such genuine quantum effects have been observed in a variety of systems, in the so-called strong-coupling regime of cavity QED, when the atom-cavity coupling strength is larger than any dissipation rate~\cite{Rempe:1987, Reithmaier:2004, Wallraff:2004, Peter:2005}.

Recently, experimental progress in tailoring the light-matter interaction has made it possible to achieve a coupling strength that is comparable or even larger than the cavity frequency $\omega_c$~\cite{Devoret:2007, Bourassa:2009, Todorov:2010, Niemczyk:2010, Forn-Diaz:2010, Nataf:2011, Forn-Diaz:2016, Yoshihara:2016,Forn-Diaz:2016b}. From a theoretical perspective, the possibility of exploring this so-called ultrastrong coupling regime has stimulated numerous studies on the quantum Rabi model that takes into account the counter-rotating terms in the atom-cavity interaction~\cite{Irish:2007, Ashhab:2010, Hwang:2010, Casanova:2010, Braak:2011,Hwang:2015, Wang:2016}. Since dissipation also plays a crucial role in most quantum optical setups, a meaningful description in this context involves a driven-dissipative scenario~\cite{Ciuti:2006,DeLiberato:2009,Beaudoin:2011,Ridolfo:2012, Henriet:2014}, in which the interplay between cavity losses and the external field drives the system into a steady state.

In such a driven-dissipative setting of the Rabi model, it has been shown recently in Ref.~\cite{LeBoite:2016} that as the coupling strength increases from $0.1\omega_c$ to $3\omega_c$, a series of transitions occurs in the output photon statistics, leading to a breakdown and revival of the so-called photon blockade effect and to a reversion to non-interacting photons. It demonstrates that the intricate interplay among the ultrastrong light-matter coupling, the external coherent driving and the dissipation stabilizes the system into a steady state exhibiting a rich quantum optical phenomenology. In this paper, going beyond the study of steady-state properties, we investigate the transient dynamics of the driven-dissipative Rabi model and show that it exhibits metastability in the ultrastrong coupling regime. Namely, we find that the convergence to the steady state is governed by a time scale significantly larger than the decay times of the atom and the cavity, giving rise to long-lived metastable states.

When the atom-cavity coupling is much smaller than the cavity frequency, the time dependency of the Liouvillian can be eliminated by a change of reference frame~\cite{Walls:2007}. All the information on the dynamics  and metastable states is then encoded in the eigenvalues and eigenfunctions of the time-independent Liouvillian \cite{Risken:1987, Vogel:1988, Risken:1988, Vogel:1989, Casteels:2016, Macieszczak:2016}. The break-down of the rotating-wave approximation in the ultrastrong coupling do not allow for such a simple transformation and the master equation remains time-dependent~\cite{Ridolfo:2012, LeBoite:2016}. To circumvent this issue we employ a Floquet-Liouville approach~\cite{Ho:1986,Grifoni:1998}: By applying Floquet theory to the Linblad master equation we reduce the time-dependent master equation to a time-independent eigenvalue problem in an enlarged Hilbert space.
 
Within this theoretical framework, we compute the long-time dynamics in the weak-excitation regime, for a driving field resonant with the second available transition. We find that the corresponding Liouvillian gap becomes significantly smaller than the natural decay rates as one increases the atom-cavity coupling strength and relate this feature to the dressed-state properties of the Rabi Hamiltonian. More specifically, a central role is played by a parity shift occurring in the spectrum, resulting in the existence of two distinct decay channels. Metastability stems from the interplay between the two different time scales involved in these two channels. The Floquet-Liouville formalism also allows us to derive analytical expressions for the set of all possible metastable sates in terms of eigenvectors of the Floquet-Liouvillian and set bounds on the deviations from the steady state.
Finally, we discuss practical implications of our analysis for future experiments probing the steady-state properties of the driven-dissipative Rabi model.

The paper is organized as follows: The model is introduced in Sec. \ref{sec:model}. The first numerical evidence of a separation of time scales in the dynamics and the emergence of metastable states are presented in Sec. \ref{sec:longtime}. Section \ref{sec:floquet} is devoted to the Floquet-Liouville formalism which is applied in Sec.~\ref{sec:meta} to a more thorough and systematic analysis of metastability. In Sec. \ref{sec:noise} we evaluate the robustness of our findings when pure dephasing noise is included in the model and we conclude in Sec.~\ref{sec:conclu}. More details on Floquet theory are presented in Appendix \ref{app:floquet} and the proofs of some spectral properties of the Floquet-Liouville operator are provided in Appendix \ref{app:meta}.
\section{The model}
\label{sec:model}

We consider a single cavity mode coupled to a two-level atom described by the Rabi Hamiltonian,
\begin{equation}
\label{Hamilto_r}
H_r = \omega_c a^{\dagger}a + \omega_a\sigma_+\sigma_- -g(a+a^{\dagger})\sigma_x,
\end{equation}
where we have introduced the photon annihilation operator $a$, and the Pauli matrices $\sigma_x$, $\sigma_y$ (with $\sigma_{\pm}  = \frac{1}{2}(\sigma_x \pm i\sigma_y)$). Here, $\omega_c$ is the cavity frequency, $\omega_a$ the atomic transition frequency, and $g$ the atom-cavity coupling strength. In the following we will focus on a resonant case, i.e., $\omega_c = \omega_a$. Note that there is no general explicit expression for the eigenstates and eigenvalues of the Rabi model. In the following, it will be convenient to label them by using an important symmetry property of the Hamiltonian, namely that the parity of the total number of excitations, $\Pi=\exp[i\pi(a^\dagger a +\sigma_+\sigma_-)]$, is a conserved quantity. We will denote by $|\Psi_j ^{\pm}\rangle$ the $j^{th}$ eigenstate ($j=0,1,..$) of the $\pm$ parity subspace and by $E_{j}^{\pm}$ the corresponding energy. With these notations, the ground state of $H_r$ is the state $|\Psi_0^{+}\rangle$, which is the lowest energy state of the $+$ parity subspace;  while the first excited state of $H_r$, which corresponds to the lowest energy state of the $-$ parity subspace, is $|\Psi_0^{-}\rangle$.

We focus in this paper on a driven-dissipative scenario where the cavity is driven by a monochromatic coherent field and both the cavity and the atom are coupled to their environments, leading to dissipation. The total time-dependent Hamiltonian of the system is
\begin{equation}\label{Hamilto}
H(t) = H_r + F\cos(\omega_dt)(a+a^{\dagger}),
\end{equation}
where $F$ is the intensity of the driving field and $\omega_d$ its frequency.
The time evolution of the density matrix $\rho(t)$ is  governed by a master equation of the form,
\begin{equation}
\label{ME}
\partial_t \rho = i[\rho,H(t)] +\mathcal{L}_a\rho+\mathcal{L}_{\sigma}\rho,
\end{equation}
where the term $\mathcal{L}_a\rho+\mathcal{L}_{\sigma}\rho$ describes the dissipation of the system excitations into the  environment. In the ultrastrong coupling regime, it is crucial to take fully into account the coupling between the atom and the cavity in the derivation of the master equation~\cite{DeLiberato:2009, Beaudoin:2011}. In particular,  the atom and the cavity can no longer be regarded as being independently coupled to their own environment and the jump operators must involve transitions between eigenstates of the total atom-cavity Hamiltonian~\cite{Beaudoin:2011}. A natural basis to express the correct master equation is therefore the dressed-state basis $\{|\Psi_j^{p}\rangle \}$ with $p=\pm$, in which the Hamiltonian (without driving) is diagonal. In this basis, the dissipative part reads,
\begin{align}\label{MEdiss}
\mathcal{L}_a\rho+\mathcal{L}_{\sigma}\rho=\sum_{p=\pm}\sum_{k,j}\Theta(\Delta_{jk}^{p\bar{p}})\left(\Gamma_{jk}^{p\bar{p}}+K_{jk}^{p\bar{p}}\right)\mathcal{D}[|\Psi_j^{p}\rangle\langle \Psi_k^{\bar{p}}|],
\end{align}
where $\Theta(x)$ is a step function, i.e., $\Theta(x)=0$ for $x\leq0$ and $\Theta(x)=1$ for $x>0$, and $\bar{p}=-p$.
We have also introduced the following notation, $\mathcal{D}[\mathcal{O}] = \mathcal{O}\rho\mathcal{O}^{\dagger} - \frac{1}{2}(\rho\mathcal{O}^{\dagger}\mathcal{O} + \mathcal{O}^{\dagger}\mathcal{O}\rho)$.
The quantities $\Gamma_{jk}^{p\bar{p}}$ and $K_{jk}^{p\bar{p}}$ denote the rates of transition from a dressed-state $|\Psi_k^{\bar{p}}\rangle$ to $|\Psi_j^{p}\rangle$ due to the atomic and cavity decay, respectively; the transition rates are defined as \cite{Beaudoin:2011, Ridolfo:2012}

\begin{align}\label{trans_rates}
\Gamma_{jk}^{p\bar p} &= \gamma\frac{\Delta_{jk}^{p \bar p}}{\omega_c}|\langle \Psi_j^{p}|(a - a^{\dagger})|\Psi_k^{\bar p}\rangle|^2, \nonumber\\
 K_{jk}^{p\bar p} &= \kappa\frac{\Delta_{jk}^{p\bar p}}{\omega_c}|\langle \Psi_j^{p}|(\sigma_- - \sigma_+)|\Psi_k^{\bar p}\rangle|^2,
\end{align}
where $\Delta_{jk}^{p\bar p} = E_k^{\bar p} - E_j^{p}$ is the transition frequency and $\gamma$, $\kappa$ are respectively the cavity and the atom decay rates. Note that the transition between states belonging to the same parity space is forbidden because both operators $a - a^{\dagger}$ and $\sigma^- - \sigma^+$ change the parity of the state. In Eqs~(\ref{MEdiss}) and (\ref{trans_rates}), the usual quantum optical master equation in which the jump operators are simply $a$ and $\sigma^-$ is recovered when the coupling strength is much smaller than the cavity frequency. 

In the following, we will be interested in the long time dynamics of Eq.~(\ref{ME}).
As in most quantum optical setups, the relevant observables to characterize the system are correlation functions of the output field. As shown in Ref.~\cite{Ridolfo:2012}, the output field in the ultrastrong coupling is proportional to an operator $\dot X^+$, defined in the dressed-state basis as:
\begin{equation}
\dot{X}^+  = \sum_{p=\pm}\sum_{k,j}\Theta(\Delta_{jk}^{p\bar{p}}) \Delta_{jk}^{p\bar{p}}|\Psi_j^{p}\rangle \langle \Psi_j^{p}|i(a^\dagger-a)|\Psi_k^{\bar{p}}\rangle \langle \Psi_k^{\bar{p}}|.
\end{equation}
The two main correlation functions that we will consider are the intensity of the emitted photons, which is proportional to $I_{out}=\langle \dot X^-\dot X^+\rangle$, and the second-order correlation function, which reads
\begin{equation}
g^{(2)}(0) = \frac{\langle \dot X^-\dot X^-\dot X^+ \dot X^+\rangle}{\langle \dot X^-\dot X^+\rangle^2}.
\end{equation}

Note that except for a sufficiently small $g$, where the rotating approximation on qubit-cavity coupling can be applied, Eq.~(\ref{ME}) generally does not have a particular rotating-frame where the equation becomes time-independent. Therefore, the solution has a residual oscillation at the driving frequency $\omega_d$ even in the $t\rightarrow\infty$ limit. The steady-state properties are then obtained by averaging the solution over several driving periods, which corresponds to a time integrated measurement in an actual experiment~\cite{Ridolfo:2012}.
%
%
\section{Long time dynamics and separation of time scales}
\label{sec:longtime}
\begin{figure}[h]
 	\centering
	\includegraphics[width = 0.49\columnwidth]{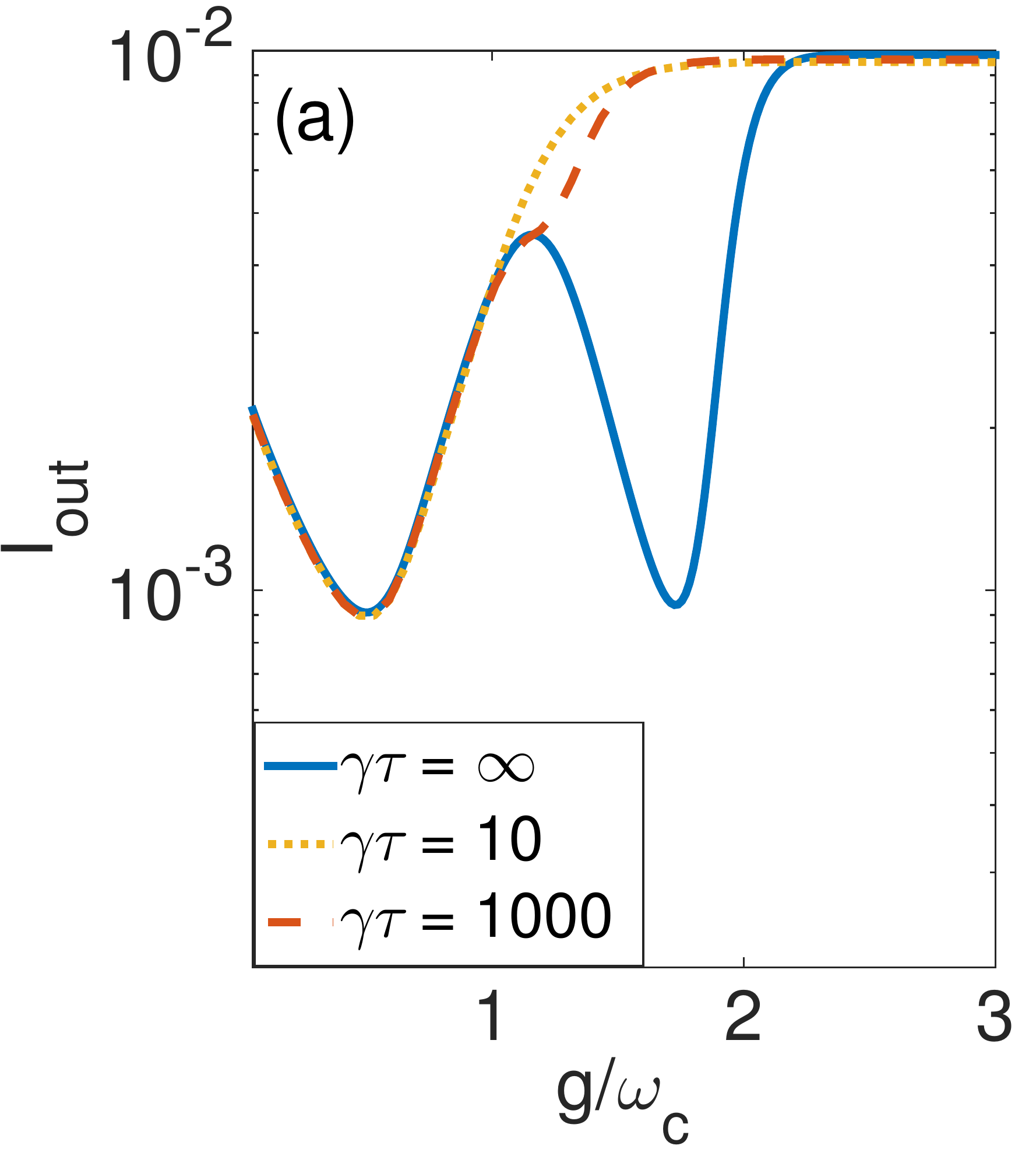}
	\includegraphics[width = 0.49\columnwidth]{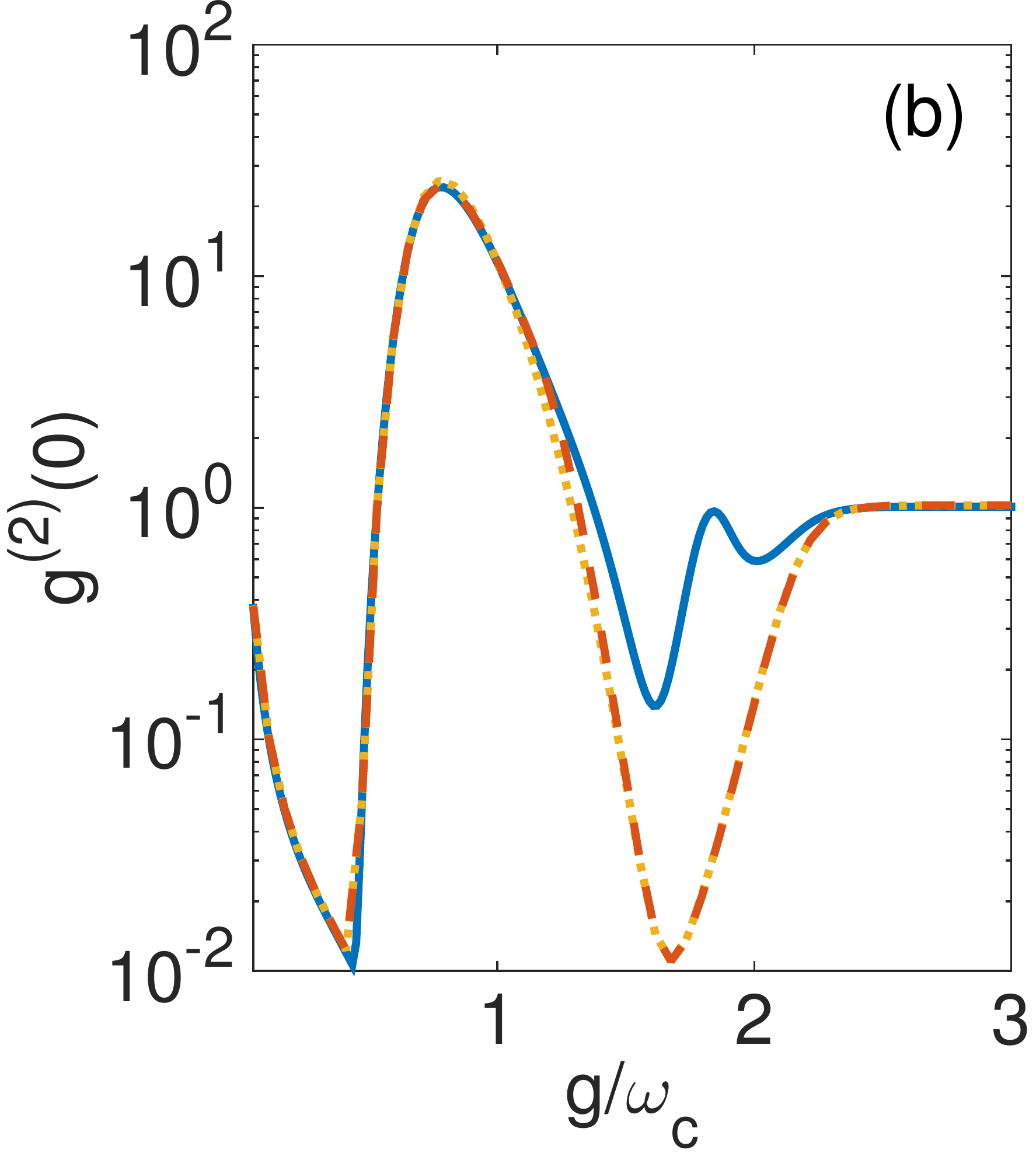}
\caption{ (a) Output intensity $I_{\mathrm{out}}$ and (b) second-order correlation function $g^{(2)}(0)$ as a function of $g/\omega_c$ for different simulation times $\tau$. The external driving field is resonant with the transition $|\Psi_0^+\rangle \to |\Psi_1^-\rangle $ and its intensity is $F/\gamma = 0.1$. The dissipation parameters are $\gamma = \kappa = 10^{-2}\omega_c$.}
\label{fig:timeinteg}
\end{figure}

In Ref.~\cite{LeBoite:2016} we have shown that in terms of output photon statistics, the most interesting properties are obtained when driving the second available transition, $|\Psi_0^+\rangle\to|\Psi_1^-\rangle $ (See Fig.~\ref{fig:spec}). We will therefore also focus on this driving scenario in all that follows. One of the main characteristic of the steady state is then that the $g^{(2)}(0)$ function exhibits a nonmonotonic behavior as a function of the coupling strength. More precisely, four different phases of photon emission can be identified: The photon blockade effect that is well-known to occur in the strong coupling regime [$\gamma/\omega_c,\kappa/\omega_c\ll g/\omega_c \ll 1$] persists up to a coupling strength  $g/\omega_c\sim0.45$. It is then followed by a break-down and revival of the photon blockade effect (for $0.45\lesssim g/\omega_c\lesssim1$ and $1\lesssim g/\omega_c\lesssim2.5$ respectively), and a transition to a noninteracting regime (for $ g/\omega_c \gtrsim 2.5$). These results are summarized in Fig.~\ref{fig:timeinteg}, where the blue solid line shows the output intensity $I_{out}$ and $g^{(2)}(0)$ in the steady state as a function of the coupling strength $g/\omega_c$. The intensity of the driving field and the dissipation rates are chosen such that the system stays in a weak-excitation regime: $\gamma = \kappa = 10^{-2}\omega_c$ and $F/\gamma = 0.1$. 

Figure~\ref{fig:timeinteg} also shows the same quantities obtained for long but finite simulation times $\tau$  (where the system is assumed to be in the ground-state at $t = 0$). For both finite-time and  steady-state values, fast oscillations are eliminated by averaging over one period of the driving frequency (a time much smaller than the decay time)~\cite{Ridolfo:2012,LeBoite:2016}. Surprisingly, we observe that the long-time dynamics in the regime where the revival of the photon-blockade occurs, i.e., $1\lesssim g/\omega_c\lesssim2.5$, sharply stands out from other coupling strengths: The output intensity and the correlation function are far from having reached their steady-state values even after a time significantly longer than the natural relaxation time, i.e.,  $\tau = 1000/\gamma$, while for both $g<1$ and $g>2.5$ the steady-state values are already reached for $\gamma\tau = 10$.

\begin{figure}
 	\centering
	\includegraphics[width = 0.98\columnwidth]{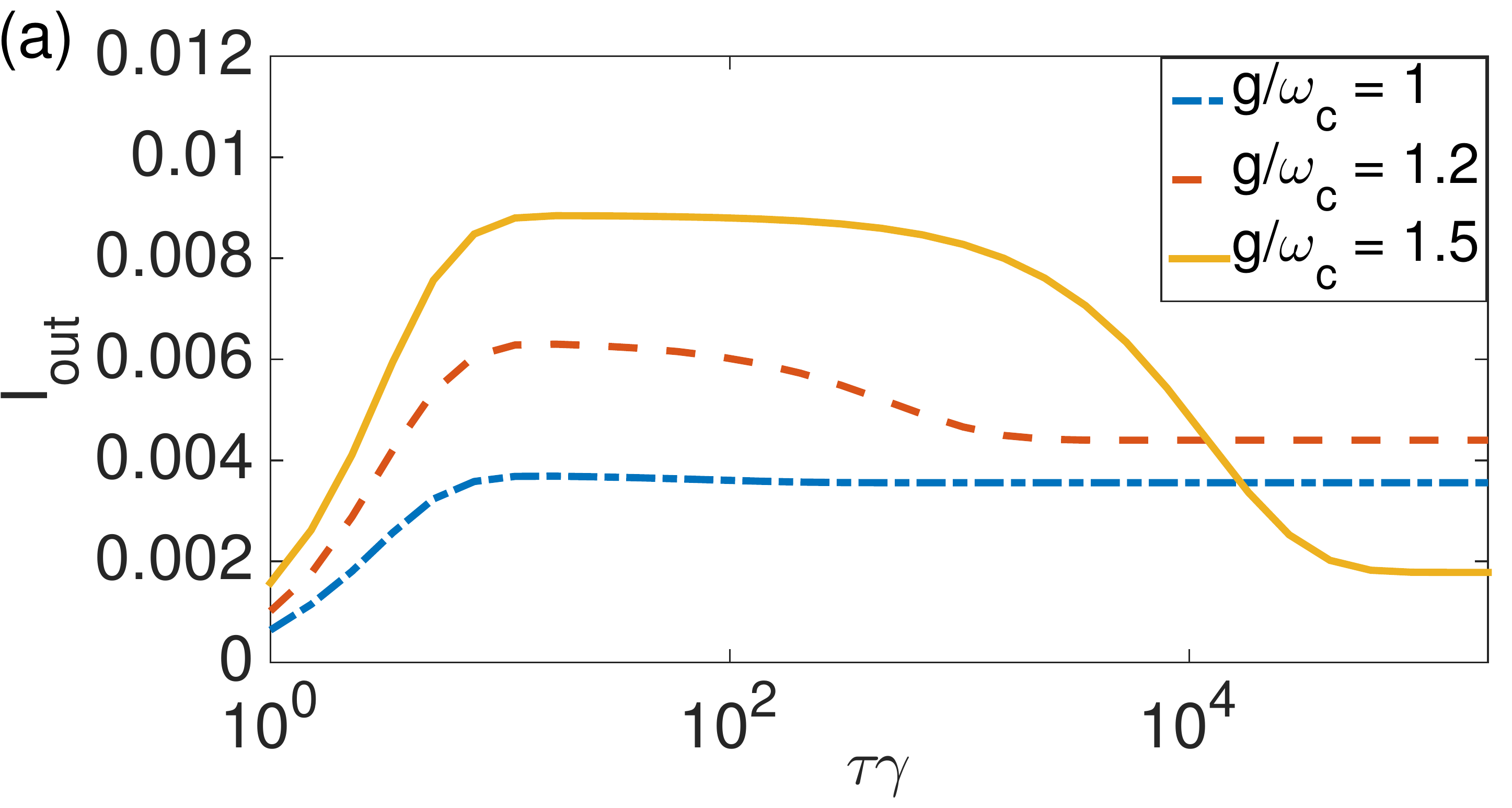}\\
	\includegraphics[width = 0.98\columnwidth]{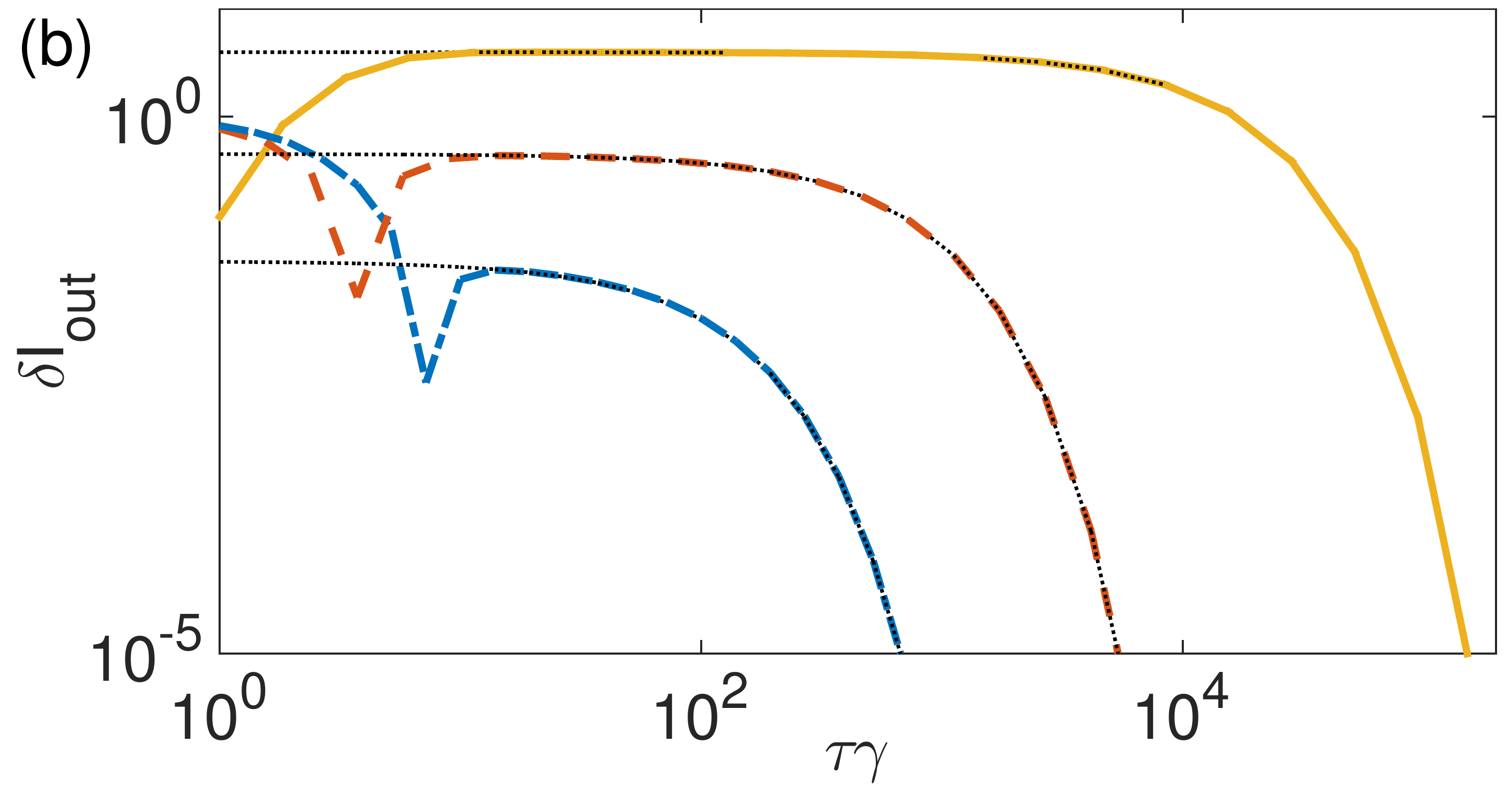}
\caption{ Long-time dynamics.  (a) $I_{out}$ as a function of time for different values of the coupling strength. For each point the result is obtained by averaging over one period (of the driving field). There is a clear emergence of metastability as $g$ is increased. (b)The quantity $\delta I_{out} = |I_{out}(\tau)-I_{out}(\infty)|/I_{out}(\infty)$ as a function of time. The same averaging procedure over one period is applied. Black dotted lines are exponential fits $\propto e^{\Omega_{0,1}\tau}$, where $\Omega_{0,1}$ is the non-zero eigenvalue of the Floquet-Liouvillian with the smallest absolute real part.}
\label{fig:timedyn}
\end{figure}

These unexpected, large discrepancies between the exact steady-state values and the finite-time simulations in the ultrastrong coupling regime suggest the emergence of a new relaxation time scale. To explore this further, we compute numerically the exact long-time dynamics of the output intensity for different values of $g/\omega_c$.
In Fig.~\ref{fig:timedyn} (a) $I_{out}$ is shown as a function of time $\tau$, for times up to $\tau\gamma = 10^5$, and for $g/\omega_c = 1$, 1.2 and 1.5. The driving and dissipation parameters are the same as in Fig. \ref{fig:timeinteg}. For $g/\omega_c = 1$ (blue dashed-dotted line), there is only one time scale in the transient dynamics and the steady-state value is reached for $1< \tau\gamma < 10$. This is a common feature for any coupling strength $g<1$. For $g = 1.2$, (dashed red lines), this simple picture is significantly modified. The steady-state value is only reached for $\tau\gamma > 10^3$ and two distinct phases in the transient dynamics are visible:  a first evolution leads the system to an intermediate state for $\tau\gamma \approx 10$, followed by a slower decay to the steady state. This separation of time scales in the dynamics is greatly amplified for $g/\omega_c = 1.5$ (solid yellow line). In this case, the intermediate state is a long-lived metastable state. The output intensity is quasi-constant for a large time interval $10 \lesssim \tau\gamma \lesssim 10^3$ and reaches its asymptotic value only for $\tau\gamma \approx 10^5$. The transient dynamics is thus characterized by a gap between the two time scales for fast and slow decay processes, giving rise to metastable states.

The numerical results presented in Figs.~\ref{fig:timeinteg} and~\ref{fig:timedyn} are one of the main findings of the present paper. 
They will be of significant experimental relevance for any setup in which the time scale of the the experiment is shorter than the time necessary to reach the steady state. In this case, the measured properties of the system in the long time limit would be that of metastable states and not of the true steady state.

The principal aim of the remaining part of the paper is to establish a proper understanding of our numerical observations and explore the metastability in the driven-dissipative Rabi model in a systematic fashion. In the case of \emph{time-indepedent} master equation, the time-scale of the transient dynamics and the properties of the metastable states can be understood in terms of spectral properties of the Liouvillian governing the time-evolution~\cite{Macieszczak:2016}. To tap into this existing framework and investigate metastability in our \emph{time-dependent} setting, the master equation in Eq.~(\ref{ME}) should therefore be cast into a time-independent form. However, due to the presence of the counter-rotating terms, there does not exist a reference frame where the time dependency is eliminated. Instead, as we will see in the next section, a time-independent formulation can be established by employing a Floquet-Liouville approach~\cite{Ho:1986}.

In this framework, eigenvalues of a Floquet-Liouvillian operator will play the same role as those of the usual Liouvillian. To illustrate this idea and motivate further the use of Floquet theory, we anticipate on what will follow and show on Fig.~\ref{fig:timedyn}~(b) the quantity $\delta I_{out} = |I_{out}(\tau)-I_{out}(\infty)|/I_{out}(\infty)$ as a function of time. The different values of the coupling strength and the other parameters correspond to that of Fig.~\ref{fig:timedyn}~(a). For each values of the $g/\omega_c$, the black dotted lines show an exponential fit with the corresponding eigenvalue $\Omega$ of the Floquet-Liouvillian operator, which will be introduced in the following section. The perfect agreement in the long-time limit is consistent with the separation of time scale described previously; after a sufficiently long time, only one slow-decaying component remains.

%
%
\section{ Floquet-Liouville approach}
\label{sec:floquet}
Floquet theory applies to linear differential equations with periodic coefficients~\cite{Floquet:1883} and, in the present context, can be used to reduce the time-dependent master equation to a time-independent eigenvalue problem in an enlarged Hilbert space. Although this so-called Floquet-Liouville approach is known and has found applications in various fields \cite{Grifoni:1998, Chu:2004}, it has not, to the best of our knowledge, been directly applied to the current setting of the driven and dissipative Rabi model. 
We therefore find it useful to present in this section the general formalism that lies at the core of our analysis. Further details on Floquet theory have also been included in Appendix \ref{app:floquet}. As a useful comparison we refer to Ref. \cite{Hausinger:2011} where Floquet theory is applied to a \emph{closed} Rabi model under strong driving.

The master equation given in Eq.~(\ref{ME}) can be written as
\begin{equation}\label{MEshort}
\partial_t \rho = \mathscr{L}(t) \rho,
\end{equation}
where $\mathscr{L}$ is a periodic linear superoperator acting on the density matrix $\rho$ and satisfying $\mathscr{L}(t+T)=\mathscr{L}(t)$, where $T = 2\pi/\omega_d$. In the following we will denote by $\mathcal{H}$ the Hilbert space of the system. ($\rho$ is then an element of $\mathcal{H}^2$.)

The Floquet theorem states that there exist solutions of Eq.~(\ref{MEshort}) of the form
\begin{equation}\label{floquetSol}
\rho(t) = \sum_{\alpha}c_{\alpha}e^{\Omega_{\alpha}t}R_{\alpha}(t).
\end{equation}
Here, $R_{\alpha}(t)$ is a periodic function of period $T$ and $\Omega_{\alpha}$ is a complex number, which are eigenfunctions and eigenvalues, respectively, of the following operator
\begin{equation}\label{eqPeriodic_main}
(\mathscr{L}(t)-\partial_t)R_{\alpha}(t) = \Omega_{\alpha}R_{\alpha}(t).
\end{equation}
Note that this last equation does not define a unique set of eigenvalues and eigenfunctions $\{\Omega_\alpha,R_\alpha\}$, the following transformation
\begin{align}
\Omega_{\alpha} &\to \Omega_{\alpha} -ik\omega_d, \\
R_{\alpha}(t) &\to e^{ik\omega_d} R_{\alpha}(t) \label{relEig},
\end{align}
with $k\in \mathbb{Z}$, gives exactly the same solution for $\rho(t)$. In the remainder of this section we will therefore label the eigenvalues and eigenfunctions with two indices $\alpha$ and $k$, the sets $\{\Omega_{\alpha,0}R_{\alpha,0}\}$ and $\{\Omega_{\alpha,k},R_{\alpha,k}\}$ being linked by the above transformation.

The key element in Eq.~(\ref{eqPeriodic_main}) is that all the functions appearing in it are periodic. The problem can therefore be made time-independent by applying a Fourier transform. Equation \eqref{eqPeriodic_main} becomes
\begin{equation}\label{floquetEig}
\sum_{m = -\infty}^{\infty} \mathscr{L}^{(n-m)}R^{(m)}_{\alpha,k} +in\omega_dR_{\alpha,k}^{(n)} = \Omega_{\alpha}R_{\alpha,k}^{(n)},
\end{equation}
where we have used the following convention for the Fourier series, $R_{\alpha,k}(t) = \sum_{n = -\infty}^{\infty} R_{\alpha,k}^{(n)}e^{-i n\omega_d t}$, $\mathscr{L}(t) = \sum_{n = -\infty}^{\infty} \mathscr{L}^{(n)}e^{-i n\omega_d t}$.

Equation~(\ref{floquetEig}) is an eigenvalue problem in an enlarged Hilbert space and is sufficient, in this formulation, to find the expression of $\rho(t)$. For practical purposes, it is useful to go one step further and make the structure of the enlarged Hilbert space more explicit. This Hilbert space, sometimes called Floquet space is the space of $T$-periodic matrices on $\mathcal{H}^2$. Formally, it is the tensor product $\mathcal{H}^2\otimes \mathcal{T}$, where $\mathcal{T}$ denotes the Hilbert space of $T$-periodic functions. 

As a basis for the space $\mathcal{T}$, a natural choice is obviously the functions $\phi_n(t) = e^{-in\omega_dt}$. Following Refs.~\cite{Grifoni:1998,Hausinger:2010}, we will denote $\phi_n$ by $|n)$ and write $\phi_n(t) = (t|n)$. With these notations, we represent the periodic matrix $R_{\alpha,k}(t)$ by a vector $|R_{\alpha,k}\rangle \rangle$  in $\mathcal{H}^2\otimes\mathcal{T}$ , defined as
 \begin{equation}\label{FloNot}
|R_{\alpha,k}\rangle \rangle = \sum_{n= -\infty}^{\infty} R_{\alpha,k}^{(n)}\otimes|n),
\end{equation}
and we have $R_{\alpha,k}(t)=(t|R_{\alpha,k}\rangle \rangle$ by definition. This equation can therefore be seen as another way of writing the Fourier series of a periodic function. Within this framework, the eigenvalue problem of Eq.~(\ref{floquetEig}), can be written as
\begin{equation}\label{eigProbFlo}
\tilde{\mathscr{L}}|R_{\alpha,k}\rangle \rangle = \Omega_{\alpha,k}|R_{\alpha,k}\rangle \rangle.
\end{equation}
where the operator $\tilde{\mathscr{L}}$ acts on element of $\mathcal{H}^2\otimes \mathcal{T}$.
%
As $\tilde{\mathscr{L}}$ is not Hermitian, it is necessary to distinguish the right eigenvectors defined above from the left eigenvectors obeying
\begin{align}
\tilde{\mathscr{L}}^{\dagger}|L_{\alpha , k} \rangle \rangle &= \Omega^*_{\alpha , k}|L_{\alpha , k} \rangle \rangle.
\end{align}
We also introduce a scalar product on $\mathcal{H}^2\otimes \mathcal{T}$,
\begin{equation}\label{scalarProd}
\langle \langle A|B\rangle\rangle = \sum_n \mathrm{Tr}[A^{(n) \dagger}B^{(n)}],
\end{equation}
which derives from the usual scalar product on $\mathcal{T}$, $(f|g) = \frac{1}{T}\int_{0}^T f^*(t)g(t)\mathrm{d}t$ and the scalar product on $\mathcal{H}^2$, $\langle A|B\rangle = \mathrm{Tr[A^{\dagger}B}]$.

Putting all this together, we can finally express the time evolution of the density matrix, i.e., the solution of Eq.~(\ref{MEshort}), in terms of the eigenvalues and the left and right eigenfunctions of the Floquet-Liouville operator $\tilde{\mathscr{L}}$. The first step is to express an initial density matrix of the system $\rho_0$ in Floquet space, e.g., $|\rho_0\rangle\rangle =\rho_0\otimes|0)$, and then decompose it in terms of eigenfunctions of $\tilde{\mathscr{L}}$,
\begin{equation}
|\rho_0\rangle\rangle = \sum_{\alpha,k}c_{\alpha,k}|R_{\alpha,k}\rangle \rangle,
\end{equation}
with $c_{\alpha,k} =  \langle\langle L_{\alpha,k}|\rho_0\rangle\rangle$. Note that for a given initial density matrix $\rho_0$, the choice of the $|\rho_0\rangle\rangle$ is not unique, but this arbitrariness has no influence on the dynamics (see Appendix \ref{app:floquet} for a proof of this statement).

The time-evolution of this initial state then immediately follows as
\begin{equation}
|\rho(t)\rangle\rangle = \sum_{\alpha,k}c_{\alpha,k}e^{\Omega_{\alpha,k}t}|R_{\alpha,k}\rangle \rangle,
\end{equation}
which is the solution of Eq.~(\ref{MEshort}) expressed in the Floquet space. In this expression, the non-periodic part of the dynamics  appears explicitly in $e^{\Omega_{\alpha,k}t}$, while the periodic part of the dynamics is implicitly encoded in $|R_{\alpha,k}\rangle \rangle$. As a final step, the solution can be expressed in the original Hilbert space using $\rho(t)=(t|\rho(t)\rangle\rangle$ and $R_{\alpha,k}(t)=(t|R_{\alpha,k}\rangle \rangle$, that is,
\begin{equation}\label{rhoDyn}
\rho(t) = \sum_{\alpha,k}c_{\alpha,k}e^{\Omega_{\alpha,k}t}R_{\alpha,k}(t).
\end{equation}
Note that in Eq.~(\ref{rhoDyn}), the summation is performed over both indices $\alpha$ and $k$, while the Floquet theorem as expressed in Eq.~(\ref{floquetSol}) involves only a sum over $\alpha$. The sum over $k$ can be suppressed by using Eq.~(\ref{relEig}) and writing Eq.~(\ref{rhoDyn}) in terms of eigenvalues and eigenvectors belonging only to the ``first Brillouin zone'', $\Omega_{\alpha,0}$ and $|R_{\alpha,0}\rangle\rangle$. The final expression is then strictly equivalent to Eq.~(\ref{floquetSol}) and reads
\begin{align}\label{rhoDynFin}
\rho(t) = \sum_{\alpha}c_{\alpha}e^{\Omega_{\alpha,0}t}R_{\alpha,0}(t),
\end{align}
where we have introduced the more compact notations $c_{\alpha} = \sum_n c_{\alpha,n}$.

The structure of $\mathscr{L}$ guaranties that one of the eigenvalues, e.g. $\Omega_{0,0}$, is equal to zero \cite{Ho:1986}. The other eigenvalues are complex with a negative real part that determine the different time scales of the transient dynamics. Taking the limit $t \to + \infty$ in Eq.~(\ref{rhoDynFin}), we also see that the asymptotic density matrix is periodic and given by $\rho_{\infty}(t) = c_{0}R_{0,0}(t)$. In addition, the condition $\mathrm{Tr}[\rho_{\infty}(t)] = 1$ implies that the coefficient $c_0$ does not depend on the initial state and is simply a normalization constant. Absorbing it in the definition of $R_0(t)$, we can write $\rho_{\infty}$ as
\begin{equation}\label{rhoInf}
\rho_{\infty}(t) = R_{0,0}(t).
\end{equation}
%

Equations~(\ref{rhoDynFin}) and~(\ref{rhoInf}) show that the theory presented in this section gives the appropriate framework for investigating long-time properties of the system. It provides a direct access to the time scales involved and an efficient way to compute the time evolution of $\rho(t)$ for arbitrary long times without having to perform any time integration of the master equation. In the next section, we use these results to systematically investigate the long time dynamics and metastability in the driven-dissipative Rabi model.
%
\section{ Metastable states}
\label{sec:meta}
To apply the results of the previous section to our specific setting, let us first give a more explicit expression for the Liouville-Floquet operator corresponding to Eq.~(\ref{ME}). Making use of the notation introduced in Sec. \ref{sec:model}, the matrix elements of $\rho$, are expressed in the dressed state basis as $\langle \Psi_{i}^p|\rho|\Psi_{i'}^{p'}\rangle$, where $i,i' \in \mathbb{N}$ and $p,p' \in \{\pm\}$, and are therefore labeled by a set of four indices $\{i,p,i',p'\}$. To simplify the notation in the corresponding Floquet space we will denote by a single greek letter such a set of indices. Using also the  basis $|n)$ introduced in the previous section for periodic functions, we deduce from Eq.~(\ref{floquetEig}) that the matrix elements of the Floquet-Liouville operator $\tilde{\mathscr{L}}$ read
\begin{equation}\label{floquetMatelem}
\langle \langle \eta,n|\tilde{\mathscr{L}}|\beta,m\rangle\rangle = \mathscr{L}_{\eta \beta}^{(n-m)} +in\omega_d\delta_{nm}\delta_{\eta \beta},
\end{equation}
where here $\mathscr{L} = i[\cdot, H] + \mathcal{L}_a+ \mathcal{L}_\sigma$. As in Sec. \ref{sec:floquet}, $\mathscr{L}^{(k)}$ refers to the $k$th Fourier component of $\mathscr{L}$. Note that the driving frequency appears explicitly in $\tilde{\mathscr{L}}$ in the form of a diagonal term. Moreover, since the time-dependency of the driving field is expressed through a cosine function, only matrix elements of $\tilde{\mathscr{L}}$ with $n-m = 0$ or $\pm 1$ are nonvanishing.

All the numerical results presented in this paper have been obtained by diagonalizing $\tilde{\mathscr{L}}$ as expressed in Eq.~(\ref{floquetMatelem}) and computing the dynamics through Eq. (\ref{rhoDynFin}). Within this framework, the results of Fig. \ref{fig:timedyn} are straightforward to interpret. In particular, in the long-time limit, the reported exponential decay is governed by the eigenvalue $\Omega_{\alpha,0}$  of $\tilde{\mathscr{L}}$ that satisfies $ \mathrm{Re}[\Omega_{\alpha,0}] \neq 0$ and that has the smallest absolute real part.
\begin{figure}
 	\centering
	\includegraphics[width = 0.98\columnwidth]{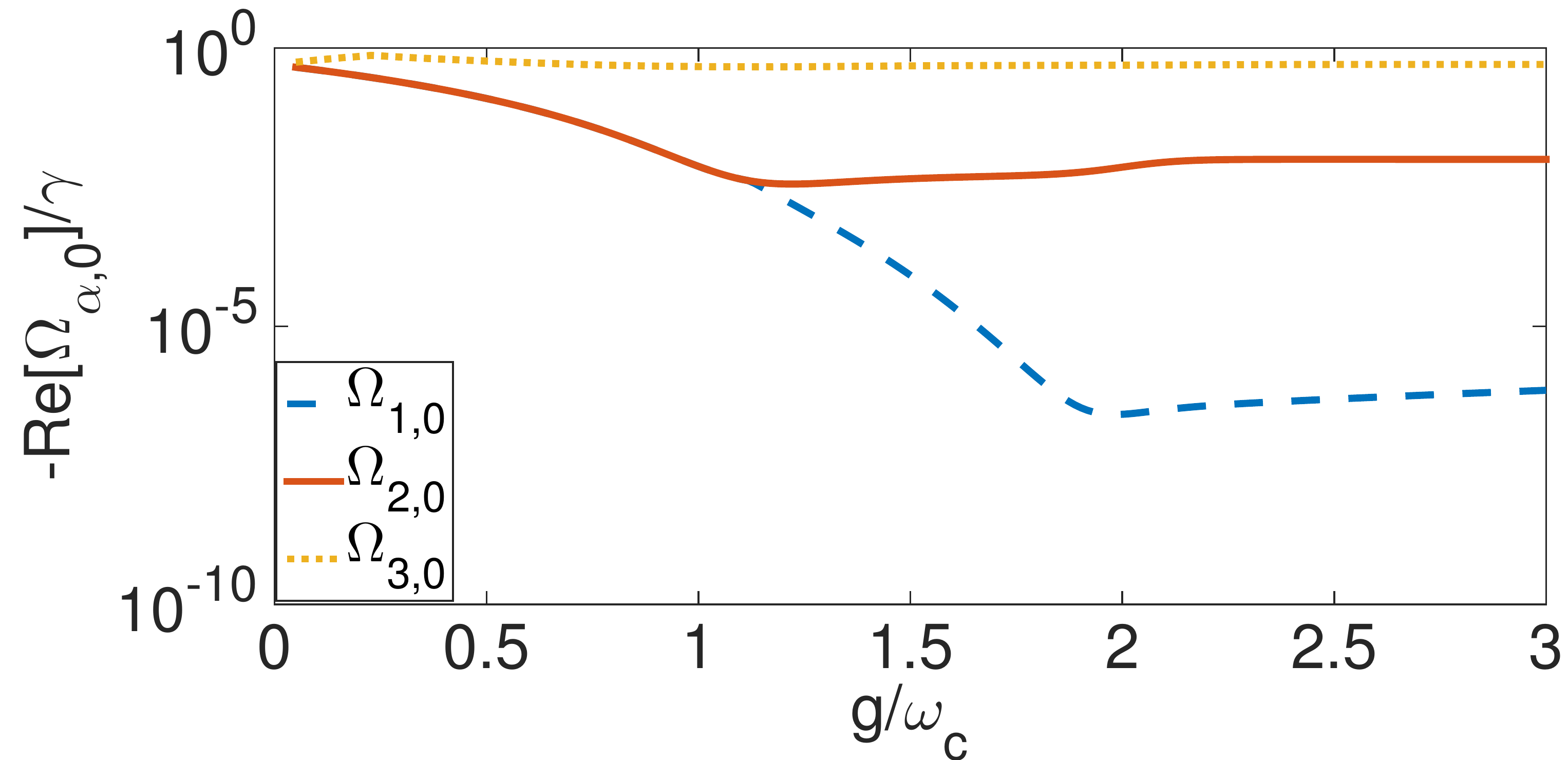}
\caption{Separation of time scales and Liouvillian gap. Real part of the first three non-zero eigenvalues, $\Omega_{1,0}$ (solid blue line), $\Omega_{2,0}$ (dashed red line) and $\Omega_{3,0}$ (yellow dotted line), as a function of $g/\omega_c$. The eigenvalues are labeled in such a way that $|\mathrm{Re}[\Omega_{\alpha,0}]|<|\mathrm{Re}\Omega_{\alpha+1,0}|$.}
\label{fig:liouvgap}
\end{figure}

More importantly,  we can now define a general criteria  for the appearance of metastability in the system: metastable states exist if there is at least one non-zero eigenvalue $\Omega_{\alpha,0}$ of $\tilde{\mathscr{L}}$ satisfying $|\mathrm{Re}[\Omega_{\alpha,0}]|\ll \gamma$. For convenience, let us label the eigenvalues of $\tilde{\mathscr{L}}$ in such a way that 
$|\mathrm{Re}[\Omega_{\alpha,0}]|<|\mathrm{Re}\Omega_{\alpha+1,0}|$.
We show in Fig.~\ref{fig:liouvgap} the real part of the first three non-zero eigenvalues, $\Omega_{1,0}$, $\Omega_{2,0}$ and $\Omega_{3,0}$  as a function of $g/\omega_c$. Remarkably, $|\mathrm{Re}[\Omega_{1,0}]|$ (blue dashed line) decreases sharply for $1\lesssim g/\omega_c \lesssim 2$, and reaches $10^{-6}\gamma$ while $|\mathrm{Re}[\Omega_{3,0}]|$ and $|\mathrm{Re}[\Omega_{2,0}]|$ rapidly saturate around $\gamma$ and $0.01\gamma$ respectively. This predicts that metastable states are likely to be observed for $g\gtrsim1$, and it is in good agreement with our previous numerical observation shown in Fig.~\ref{fig:timeinteg}

To go further, it is important to keep in mind that unlike the steady state, metastable states are not unique; the one observed in an experiment will depend on the initial state. A natural task is then to determine the set of all possible metastable states and their properties. Once again, the Floquet-Liouville formalism will prove to be the appropriate tool. Let us begin the discussion by recalling two general results that can be deduced from the structure of the master equation. These results are a generalization to Floquet-Liouville formalism of metastability theory as presented, e. g., in Ref~\cite{Macieszczak:2016}.  i) If $\Omega$ is an eigenvalue of Eq.~(\ref{eqPeriodic_main}) and $R(t)$ a corresponding eigenfunction, then $R^{\dagger}(t)$ is also an eigenfunction, and the associated eigenvalue  is $\Omega^*$.
ii) If $\Omega \in \mathbb{R}$, the left and right eigenfunctions $R(t)$ and $L(t)$ can be chosen Hermitian. In terms of Fourier component, this translates into $R^{(-n)} = R^{(n)\dagger}$. Proofs of these results are provided in Appendix \ref{app:meta}. A first consequence is that the matrix $R_{0,0}$ appearing in Eq.~(\ref{rhoInf}) is Hermitian.

\begin{figure}[t]
 	\centering
	\includegraphics[width = 0.49\columnwidth]{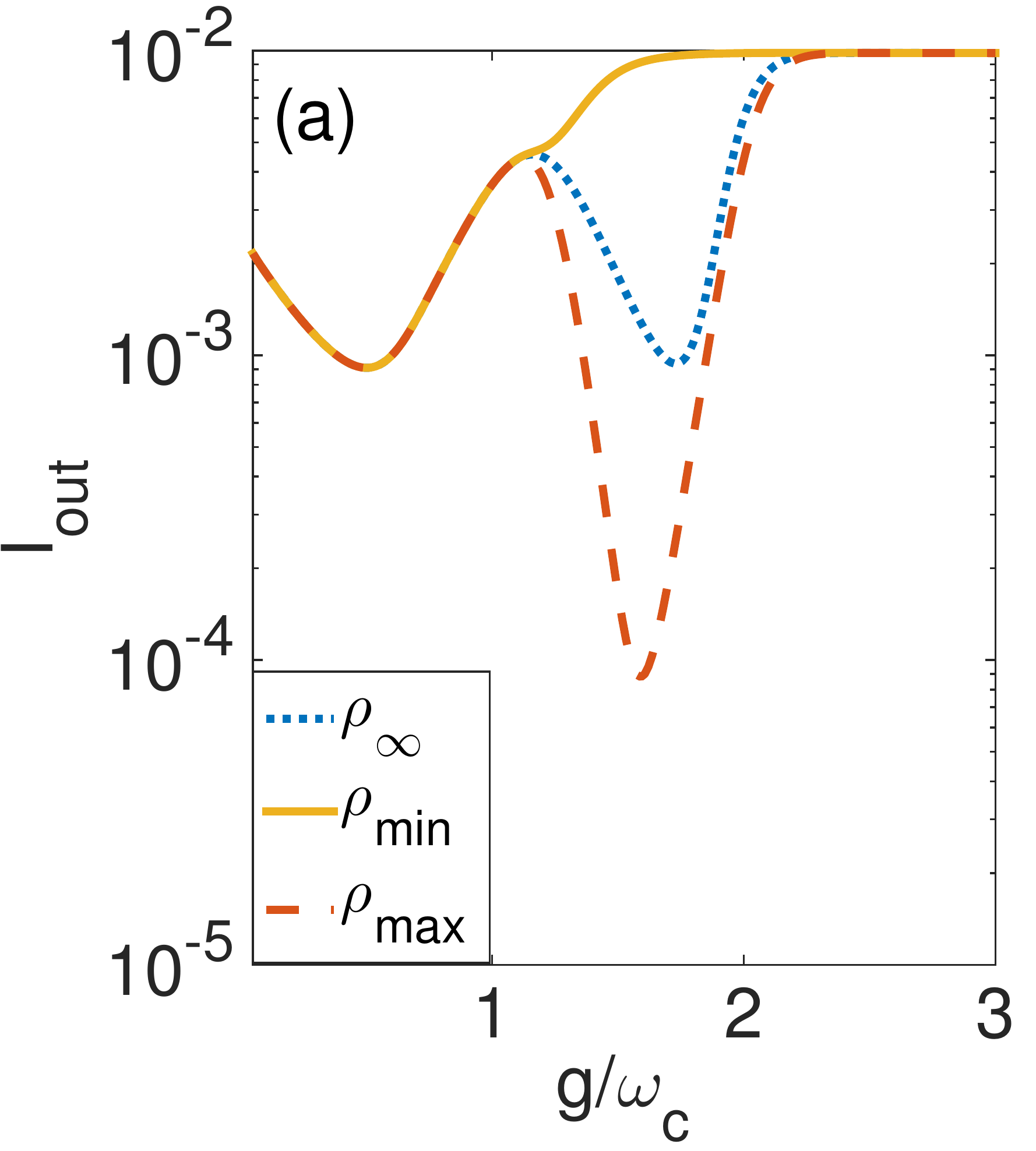}
	\includegraphics[width = 0.49\columnwidth]{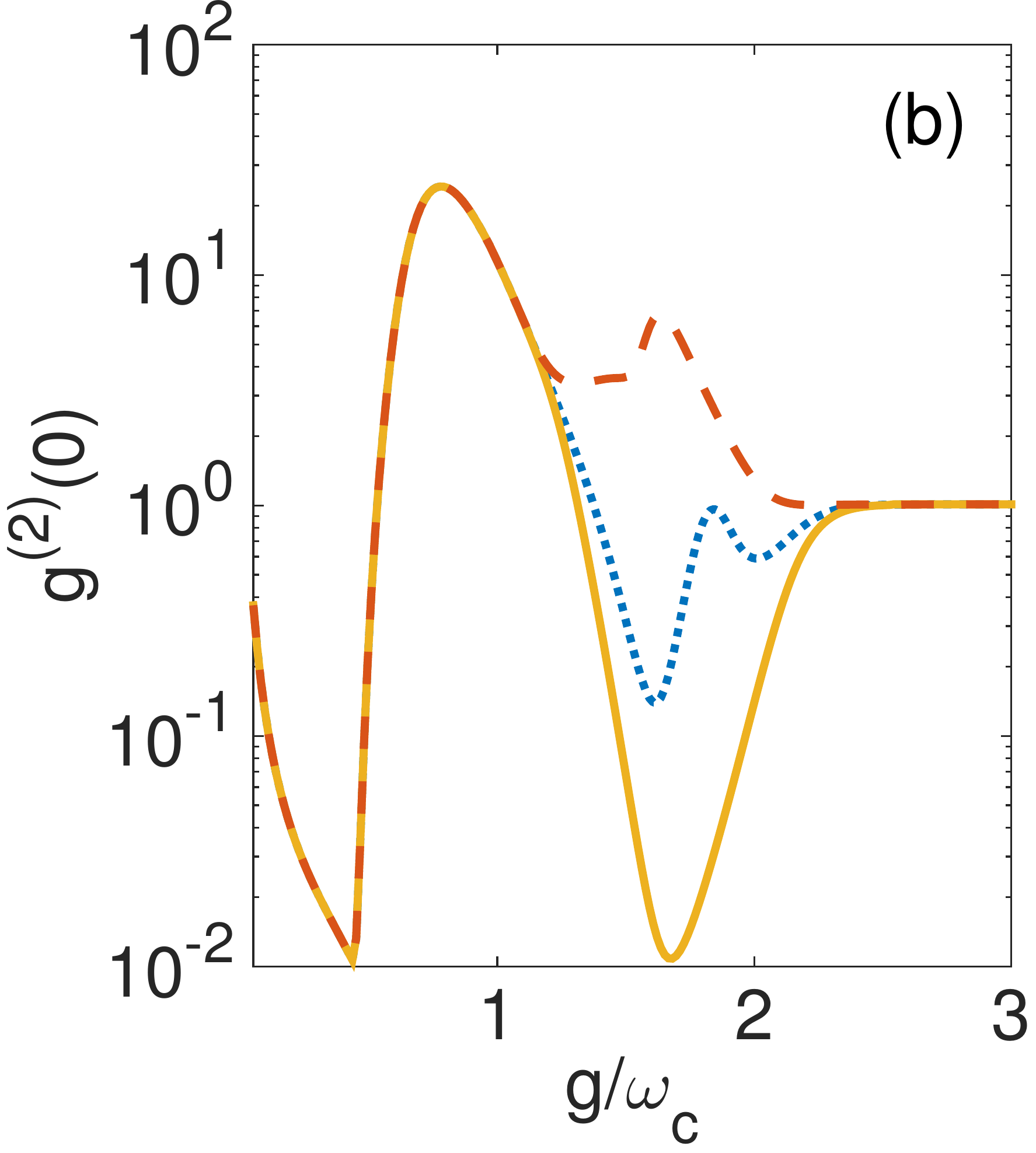}
\caption{Extremal metastable states. The output intensity and $g^{(2)}(0)$ of all possible metastable states for an arbitrary initial state lie in between the extremal metastable states values (the orange dashed line and the yellow solid line), which can be drastically different from the steady state value (blue dotted line). The observation time is set to $\tau\gamma = 10^3$.}
\label{fig:metamanifold}
\end{figure}

To find the general expression for the metastable states, we will rely on an additional property of $\Omega_{1,0}$ visible on Fig.~\ref{fig:liouvgap}: for $g/\omega_c \gtrsim 1.3$, $\Omega_{1,0}$ not only satisfies $|\mathrm{Re}[\Omega_{1,0}]|\ll \gamma$, but also  
$|\mathrm{Re}[\Omega_{1,0}]|\ll |\mathrm{Re}[\Omega_{\alpha,0}]| $ for $\alpha > 1$. 
This means that after a sufficiently long time, the density matrix will take the form
\begin{equation}\label{metaGen}
\rho(t) \approx R_{0,0}(t) + c_1R_{1,0}(t).
\end{equation}
 Another important feature of $\Omega_{1,0}$ is that it is pure real. The eigenfunction $R_{1,0}(t)$ can therefore be chosen Hermitian. Moreover, we know from Eq.~(\ref{rhoInf}) that $\mathrm{Tr}[R_{0,0}(t)] = 1$ for every time $t$, which in turn implies that $\mathrm{Tr}[R_{1,0}(t)] = 0$. Since $R_{1,0}(t)$ is Hermitian, we have also $c_1\in \mathbb{R}$.

Conversely, any matrix taking the form of Eq.~(\ref{metaGen}) with $c_1\in \mathbb{R}$ and satisfying the positivity requirement of the density matrix is a possible metastable state. In particular, the set $\mathcal{M}$ of metastable states is a convex subset of the set of density matrices $\mathcal{D}$. Furthermore, $\mathcal{M}$ is parametrized by a single real coefficient. The set of all possible values of $c_1$ is therefore a segment $[c_{\mathrm{min}},c_{\mathrm{max}}]\subset \mathbb{R}$.  

To find $c_{\mathrm{max}}$ and $c_{\mathrm{min}}$, let us go back to the general expression for the coefficients $c_{\alpha} = \sum_k c_{\alpha,k}$. Using Eq.~(\ref{scalarProd}) for the scalar product defining $c_{\alpha,k }$ and assuming that the initial state $|\rho_0\rangle\rangle$ is of the form $\rho_0 \otimes |0)$, the coefficients $c_\alpha$ can be written as
\begin{equation}
c_{\alpha} = \sum_{k} \mathrm{Tr}[L_{\alpha,k}^{(0)\dagger}\rho_0]
\end{equation}
As previously, it is more convenient to express every quantity in terms of eigenfunctions $L_{\alpha,0}$ only. It is possible through the relation $L^{(0)}_{\alpha,k} = L^{(k)}_{\alpha,0}$, which is equivalent to Eq.~(\ref{relEig}). We find
\begin{equation}
c_{\alpha} = \sum_{k} \mathrm{Tr}[L_{\alpha,0}^{(k)\dagger}\rho_0] = \mathrm{Tr}[L_{\alpha,0}(t= 0)\rho_0],
\end{equation}
where the last equality follows from the definition of $L_{\alpha,0}(t)$ and the fact that $\rho_0$ is Hermitian.
Applying this last result to $L_{1,0}$, we find that $c_{\mathrm{min}}$ is given by $c_{\mathrm{min}} = \min_{\rho \in \mathcal{D}}\mathrm{Tr[L_{1,0}(t=0)\rho]} $. A similar expression holds for $c_{\mathrm{max}}$. Given the positivity of $\rho$, the minimum is simply the smallest eigenvalue of $L_{1,0}(t=0)$ (which exist and is real since $L_{1,0}(t)$ is Hermitian). We have therefore the final result
\begin{align}
c_{\mathrm{min}} = \min \mathrm{Sp}[L_{1,0}(t=0)],\\
c_{\mathrm{max}} = \max\mathrm{Sp}[L_{1,0}(t=0)],
\end{align}
where Sp denotes the spectrum. Any metastable state will then be a convex combination of two extremal states
\begin{align}
\rho_{\mathrm{min}} = R_0(t)+ c_{\mathrm{min}}R_1(t),\\
\rho_{\mathrm{max}} = R_0(t)+ c_{\mathrm{max}}R_1(t).
\end{align}

Note that the results presented above are valid when $\Omega_{1,0}$ satisfies $|\mathrm{Re}[\Omega_{1,0}]|\ll |\mathrm{Re}[\Omega_{\alpha,0}]| $ for $\alpha > 1$. Figure~\ref{fig:liouvgap} shows that it is not the case for $g/\omega_c \sim 1$. Indeed, around this value of the coupling strength, the three eigenvalues $\Omega_{1,0}$, $\Omega_{2,0}$ and $\Omega^*_{2,0}$ are of the same order of magnitude and are all much smaller than $\gamma$. Hence, the general form of the metastable states in this regime of parameters is $\rho(t) \approx R_{0,0}(t) + c_1R_{1,0}(t)+c_2R_{2,0}(t)+ c_2^*R^{\dagger}_{2,0}(t)$. However, numerical simulations show that the eigenvalues of $L_{2,0}(t=0)$ are always much smaller than those of $L_{1,0}(t=0)$ and thus $c_{2},c_{2}^*\ll c_1$. Therefore, $R_{2,0}(t)$ and $R^{\dagger}_{2,0}(t)$ do not contribute significantly to the dynamics and the analysis of metastable states based on Eq.~(\ref{metaGen}) remains valid.

An overview of the properties of the metastable states is given in Fig.~\ref{fig:metamanifold}. The output intensity and $g^{(2)}(0)$ in $\rho_{\infty}$, $\rho_{\mathrm{min}}$ and $\rho_{\mathrm{max}}$ are plotted as a function of the coupling strength. Note that, by definition of the extremal states, all the information on the set of metastable states is contained in $\rho_{\mathrm{min}}$ and $\rho_{\mathrm{max}}$. The values shown on Fig.~\ref{fig:metamanifold} set bounds on the deviation from the true steady-state value that can be observed in an experiment. As expected from our previous results, it is for $1\lesssim g/\omega_c\lesssim 2$ that the differences between these three states in terms of observables are the highest. In particular the photon statistics differs radically, being sub-Poissonian for $\rho_{\mathrm{min}}$ and strongly super-Poissonian for $\rho_{\mathrm{max}}$.  Although metastable states also exist for higher values of $g$ ($g\gtrsim2$), the value of $I_{out} $ and $g^{(2)}(0)$ converge to the steady-state value in this case. Comparing the results of Fig.~\ref{fig:timeinteg} and Fig.~\ref{fig:metamanifold}, we find that the metastable state observed when the system is in its ground state at $t= 0$ is very close to the state $\rho_{\mathrm{min}}$. Conversly, a metastable state close to $\rho_{\mathrm{max}}$ is obtained when the initial state is the first excited state $|\Psi_0^-\rangle$ (not shown).

\begin{figure}
 	\includegraphics[width = 0.98\columnwidth]{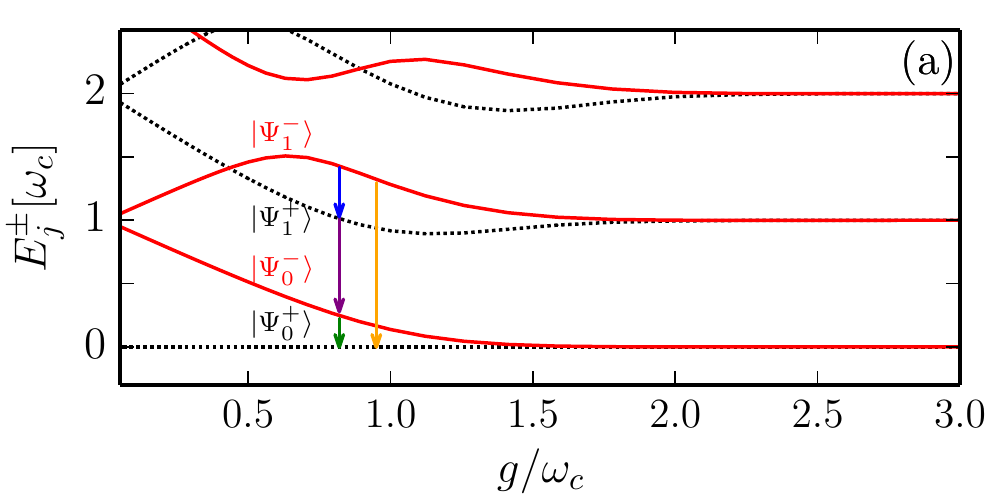}\\
	\includegraphics[width = 0.98\columnwidth]{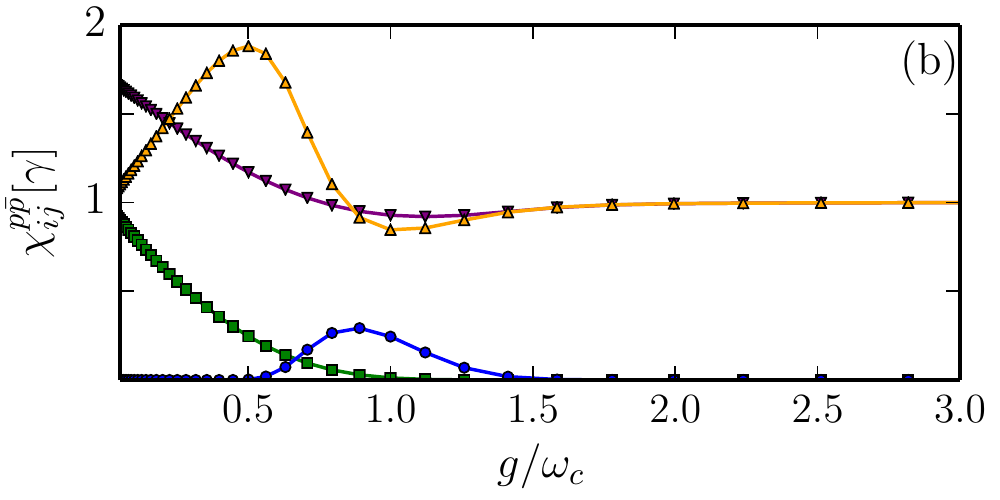}
\caption{(a) Energy spectrum of the Rabi Hamiltonian (without driving). Black dotted lines indicate energy levels with an even number of excitations while red solid lines correspond to an odd number of excitations.  Arrows show the available decay channels when driving the second transition $|\Psi_{0}^+\rangle\to|\Psi_{1}^-\rangle$. The colors match the one used in the lower panel for the transition rates. (b) Transition rates between the different dressed states, $\chi^{+-}_{00}$ (green squares), $\chi^{+-}_{11}$ (blue circles), $\chi^{-+}_{01}$ (inverted purple triangles) and $\chi^{+-}_{01}$ (yellow triangles), as a function of the coupling strength.}
\label{fig:spec}
\end{figure}
%
A qualitative explanation for the difference in photon statistics for $\rho_{\mathrm{min}}$ and $\rho_{\mathrm{max}}$ can be drawn from the dressed state properties of the Rabi model and the competing decay processes at play.  As shown in Fig.~\ref{fig:spec}(a), when the transition $|\Psi_0^+\rangle \to |\Psi_1^-\rangle $ is driven, there appear two decay channels  for $g/\omega_c \gtrsim 0.45$, after a parity shift in the spectrum has occurred~\cite{LeBoite:2016}. The first decay channel involves the direct transition $|\Psi_1^-\rangle \to |\Psi_0^+\rangle $, while the second one involves the cascaded transition $|\Psi_1^-\rangle \to |\Psi_1^+\rangle \to |\Psi_0^-\rangle \to |\Psi_0^+\rangle$. Because the direct transition leads to sub-Poissonian and the cascaded transition to super-Poissonian statistics of the output photons~\cite{LeBoite:2016}, we can expect that the competition between these two decay processes will ultimately determine the output photon statistics. 
More precisely, numerical simulations show that the metastable state $\rho_{\mathrm{min}}$ is mainly a statistical mixture of $|\Psi_0^+\rangle$ and $|\Psi_1^-\rangle$, namely $\rho_{\mathrm{min}}\approx \lambda_0|\Psi_0^+\rangle\langle\Psi_0^+| + \lambda_3|\Psi_1^-\rangle\langle\Psi_1^-|$, with $\lambda_3\ll \lambda_0$.  The metastable state $\rho_{\mathrm{max}}$ on the other hand is a statistical mixture of $|\Psi_0^-\rangle$ and $|\Psi_1^+\rangle$, $\rho_{\mathrm{max}}\approx \lambda_1|\Psi_0^-\rangle\langle\Psi_0^-| + \lambda_2|\Psi_1^+\rangle\langle\Psi_1^+|$, with $\lambda_3/\lambda_0 \approx \lambda_2/\lambda_1$. This means that $\rho_{\mathrm{min}}$ and $\rho_{\mathrm{max}}$ can be reached when the dominant relaxation process is the direct transition or the cascaded transition, respectively. Therefore, $\rho_{\mathrm{min}}$ leads to a pronounced photon blockade that can be even stronger than in the steady state while, in contrast, $\rho_{\mathrm{max}}$ shows  photon bunching (see Fig.~\ref{fig:metamanifold} (b)).

The observed metastable state depends sensitively on the initial state. For example, when the initial state is the ground state, the eigenstates $|\Psi_0^-\rangle$ and $|\Psi_1^+\rangle$ can be populated only through the cascaded transition. We show in Fig.~\ref{fig:spec}~(b) that the transition rates $\chi_{00}^{+-}$ and $\chi_{11}^{+-}$ for $|\Psi_0^-\rangle\to|\Psi_0^+\rangle$ and $|\Psi_1^-\rangle\to|\Psi_1^+\rangle$ respectively, drop sharply for $g/\omega_c\gtrsim1$, while the transition rates $\chi_{01}^{+-}$ and $\chi_{01}^{-+}$ for $|\Psi_1^-\rangle \to |\Psi_0^+\rangle$ and $|\Psi_1^+\rangle \to |\Psi_0^-\rangle$ are much higher and satisfy $\chi_{01}^{+-}\sim \chi_{01}^{-+}$. Therefore, the processes leading to the system being in the subspace $\{|\Psi_0^-\rangle, |\Psi_1^+\rangle\}$ take place at a much slower rate. Hence, on the relatively short time scale on which metastability is observed, this subspace does not play a significant role in the dynamics and the metastable state is very close to $\rho_{\mathrm{min}}$ [Fig.~\ref{fig:timeinteg} (b)].

To summarize, the general physical picture is the following: the steady state is reached when the pumping mechanisms exactly compensate the losses induced by the different decay channels. In the Rabi model, the parity shift occurring in the Hamiltonian for $g/\omega_c \approx 0.45$ leads to the existence of two distinct decay channels [Fig.~\ref{fig:spec} (a)]. Furthermore, the time scales involved in these two channels become widely separated as the coupling strength becomes larger than the cavity frequency ($g/\omega_c \gtrsim1$) [Fig.~\ref{fig:spec} (b)]. As a result, there exists an intermediate time scale in which losses from the fast decay channel are already compensated by the driving field while the other channel has not yet come into play.  In such a time interval, which is long enough to be observed experimentally, the system is in a metastable state whose properties can differ radically from those of the true steady state. Discrepancies between metastable states and the steady state are particularly sharp in the regime of coupling strength where the revival of the photon blockade takes place [$1\lesssim g/\omega_c\lesssim 2$], since in this regime the two decay channels have opposite effects on the photon statistics: the fast one favors the photon blockade effect, while the slower one destroys it by inducing additional fluctuations.

This picture however breaks down for $g\gg1$ where the energy spectrum of the Rabi model becomes quasi-linear~\cite{Hwang:2016}; in this case, the states $|\Psi_j^+\rangle$ and $|\Psi_j^-\rangle$ are quasi-degenerate and the relaxation processes also involve transitions between higher-energy states. The decay channels are now two distinct ``ladders'': $|\Psi_j^-\rangle \to |\Psi_{j-1}^+\rangle \to \dots \to |\Psi_1^-\rangle \to |\Psi_{0}^+\rangle$ when the initial state is the ground state, and $|\Psi_j^+\rangle \to |\Psi_{j-1}^-\rangle \to \dots \to |\Psi_1^+\rangle \to |\Psi_0^-\rangle$ when the system is initially in its first excited state. A separation of time scales still exists in this regime; it stems from the very low probability of transition between the two ladders through processes such as $|\Psi_{j}^+\rangle \to |\Psi_{j}^-\rangle$. However, the two channels both lead to a quasi-coherent statistics, explaining the convergence of the metastable-states properties to those of the steady state.

\section{Effect of pure dephasing noise}
\label{sec:noise}
\begin{figure}
 	\centering
	\includegraphics[width = 0.98\columnwidth]{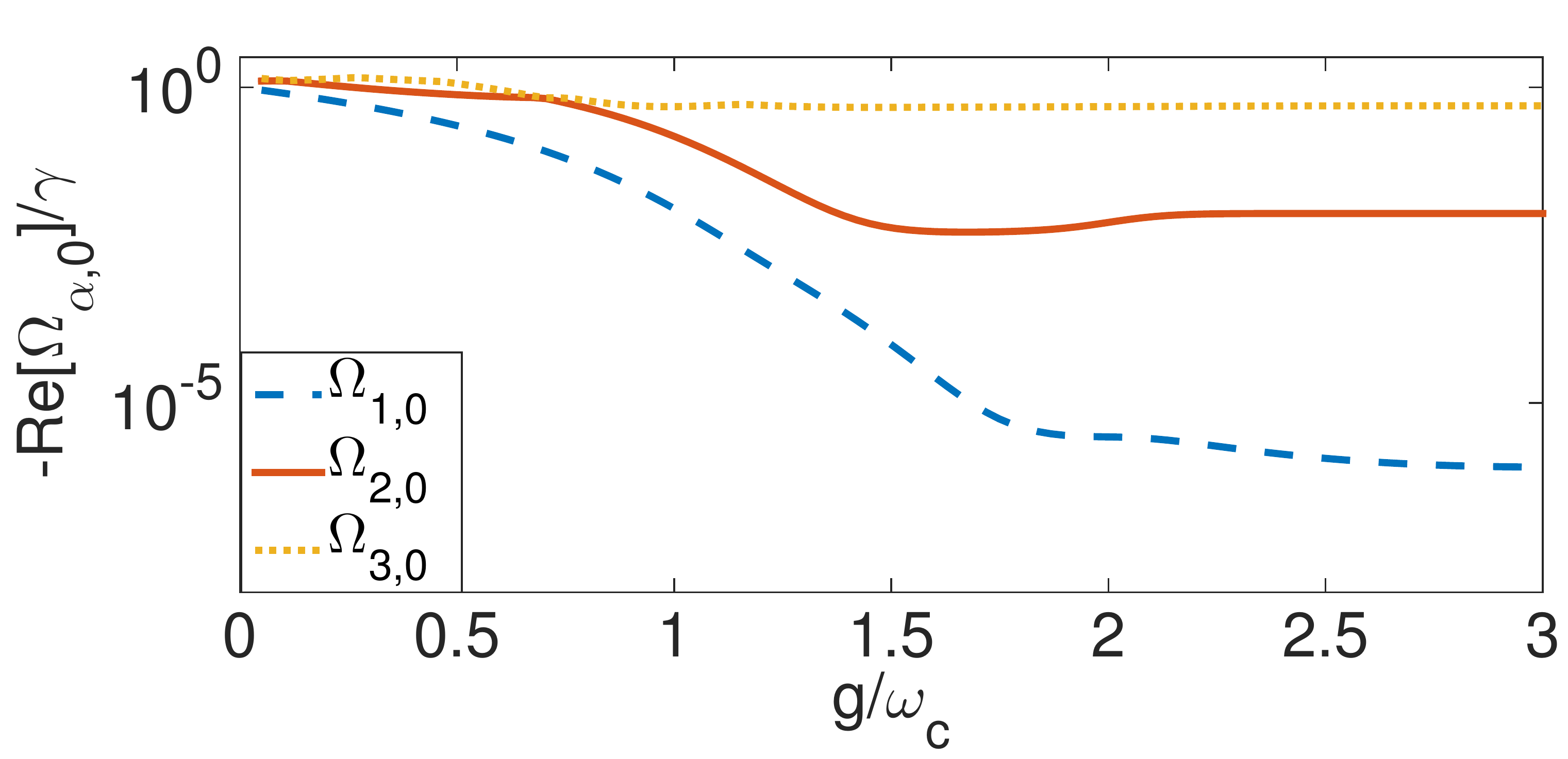}
\caption{Liouvillian gap when dephasing noise is included. Real part of the first three non-zero eigenvalues, $\Omega_{1,0}$ (solid blue line), $\Omega_{2,0}$ (dashed red line) and $\Omega_{3,0}$ (yellow dotted line), as a function of $g/\omega_c$ when dephasing noise is included. The eigenvalues are labeled in such a way that $|\mathrm{Re}[\Omega_{\alpha,0}]|<|\mathrm{Re}\Omega_{\alpha+1,0}|$. The noise parameters are $\gamma_{\phi}(\Delta_{jk}^{pp})=\gamma$.} 
\label{fig:liouvgapdeph}
\end{figure}

 In this section we evaluate the robustness of our findings against pure dephasing noise, inevitably present in any experimental setup. Following Ref.~\cite{Beaudoin:2011} we model the dephasing noise by including an additional term in the Liouvillian. Its general form is,
 \begin{align}
 \mathcal{L_{\phi}}\rho &= \mathcal{D}\left[\sum_{p=\pm}\sum_{k}\Phi^p_k|\Psi_k^p\rangle\langle \Psi_k^p|\right] \\
 &+ \sum_{p=\pm}\sum_{k,j}\Theta(\Delta_{jk}^{pp})\Phi_{jk}^{pp}\mathcal{D}[|\Psi_j^{p}\rangle\langle \Psi_k^{p}|].
 \end{align}
For this type of noise, the transition rates depend on the matrix elements of the operator $\sigma_z$ in the dressed-state basis and are given by
\begin{align}
\Phi^p_k &= \sqrt{\frac{\gamma_{\phi}(0)}{2}}\langle \Psi_k^p|\sigma_z|\Psi_k^p\rangle, \\
\Phi_{jk}^{p p} &= \frac{\gamma_{\phi}(\Delta_{jk}^{p p})}{2}|\langle \Psi_j^{p}|\sigma_z|\Psi_k^{ p}\rangle|^2.
\end{align}
These coefficients depend on the spectral density of the bath at the different transition frequencies $\Delta_{jk}^{p p}$, expressed by the function $\gamma_{\phi}(\Delta_{jk}^{p p})$. Just as in the case of the other dissipative terms, we assumed that the spectral density of the bath vanishes at negative frequency, since the system is in thermal equilibrium at zero temperature. Note that in contrast with the operators $a$ and $\sigma_-$,
the operator $\sigma_z$ can induce transitions only between states of the same parity. In principle, the additional transitions between dressed-states induced by the dephasing noise can affect the transient regime and reduce the life time of the metastable states. We show in Fig.~\ref{fig:liouvgapdeph}, numerical simulations of the Floquet-Liouvillian eigenvalues for a white dephasing noise, whose rate is comparable to the other noise sources [$\gamma_{\phi} = \gamma = \kappa$]. Globally, the real part of the eigenvalues is larger, which means that the time to reach state  is indeed reduced compared to the results of Fig.~(\ref{fig:liouvgap}). However, the clear separation of time scales is still visible and the life time of the metastable states is long enough to allow for experimental observation. Hence, there is no qualitative change and our results remain valid even when this additional noise channel is included in the model.
\section{Conclusion}
\label{sec:conclu}
In this paper, we have investigated the long-time dynamics and metastability of the driven-dissipative Rabi model in the ultrastrong coupling regime within the framework of Floquet-Liouville theory. In the ultrastrong coupling regime, the counter-rotating terms make the master equation for the driven Rabi model explicitly time-dependent, and the Floquet-Liouville theory allows one to eliminate this explicit time-dependence by considering the time evolution in an enlarged Hilbert space of periodic matrices. Our work demonstrates that the use of Floquet-Liouville theory in the driven-dissipative Rabi model not only makes an efficient calculation of arbitrarily long time-evolution possible, but also enables one to obtain analytical results and a qualitative understanding.

More specifically, we have considered a driving scenario in which the external field is resonant with the second available transition and have shown that, as the atom-cavity coupling strength becomes larger than the cavity frequency, $g/\omega_c \gtrsim 1$, the time necessary to reach the steady state becomes much larger that the natural relaxation time $1/\gamma$. Within the framework of Floquet-Liouville theory, the different time scales of the transient dynamics are understood by investigating the eigenvalues of the time-independent Floquet-Liouvillian operator. For $g/\omega_c>1$, one non-zero eigenvalue with zero imaginary part (purely dissipative mode) was found to be several orders of magnitude smaller than all the other eigenvalues, explaining the emergence of long-lived metastable states. We attributed this feature of the Floquet-Liouvillian to the existence of two decay channels for the system with different transition rates.  In particular, the transition rates for the first and third part of the cascaded transition $|\Psi_1^-\rangle \to |\Psi_1^+\rangle \to |\Psi_0^-\rangle \to |\Psi_0^+\rangle$ go to zero as $g/\omega_c$ increases. As a result, this decay channel starts to play a significant role only in the long-time dynamics. During the large time interval for which the other decay channel, $|\Psi_1^-\rangle \to |\Psi_0^+\rangle $ dominates, the system reaches a metastable state, which eventually decays into the true steady state when the second channel comes into play. 

By extending the recently developed metastability theory~\cite{Macieszczak:2016} to our time-dependent setting through the Floquet-Liouville approach, we also derived analytical expressions for the set of all possible metastable states in terms of eigenvectors of the Floquet-Liouvillian. This enabled us to set bounds on the deviation from the true steady state that could be observed in an experiment. More specifically, we showed that for $1\lesssim g/\omega_c\lesssim 2$ the photon statistics in the metastable states can differ drastically from that of the steady state ; it can either show an enhanced anti-bunching or, conversely, strong bunching. All these results were derived by considering dissipation coming from the coupling of the cavity and the atom to the environment at zero temperature. We have also performed additional simulations including pure dephasing noise and have shown that our findings remain unchanged when this etra noise channel is included in the model.

In a circuit QED experiment with a typical cavity frequency $\omega_c$ of the order of the GHz and dissipation rates similar to the one considered here [$\kappa = \gamma = 10^{-2}\omega_c$], the time scale on which metastability will be observed is of the order of 0.1 millisecond, a time sufficiently short to be reached experimentally. 

This work was supported by  the EU STREPs DIADEMS and EQUAM, the ERC Synergy Grant BioQ as well as the DFG via the SFB TRR/21 and SPP 1601.

%
%
\appendix
\section{Floquet theory and dynamics in Floquet space}
\label{app:floquet}
We give in this appendix a more detailed and self-contained presentation of Floquet theory and its formulation in the Floquet space introduced in the main text. To simplify the notations, we consider the case of a usual Schr\"odinger equation on a Hilbert space $\mathcal{H}$ of finite dimension $N$,
\begin{equation}\label{eqDiffFlo}
i\partial_t |X\rangle = A(t)|X\rangle,
\end{equation}
where $A$ is a periodic matrix of period $T$ and $X$ a vector in $\mathcal{H}$. The Floquet theorem states that there exist solutions of the form
\begin{equation}\label{floquetSolApp}
|X_{\alpha}(t)\rangle = e^{-i\epsilon_{\alpha} t} |p_{\alpha}(t)\rangle,
\end{equation}
with $|p_{\alpha}(t)\rangle$ periodic, of period $T$, and $\epsilon_{\alpha}$ a complex number.
The functions $|p_{\alpha}(t)\rangle$ are eigenfunctions of the following operator
\begin{equation}\label{eqPeriodic}
(A(t)-i\partial_t)|p_\alpha(t)\rangle = \epsilon_{\alpha}|p_{\alpha}(t)\rangle.
\end{equation}
Since all the functions that appear in Eq.\eqref{eqPeriodic} are periodic, this equation translates the original problem into an eigenvalue problem in a space of periodic functions. Let us therefore introduce the space $\mathcal{F} = \mathcal{H}\otimes T$ of periodic functions on $\mathcal{H}$. This space is a Hilbert space whose scalar product derives for the one defined on $\mathcal{H}$ and $\mathcal{T}$.
Following the notations of Ref.~\cite{Grifoni:1998,Hausinger:2010}, we define the scalar product on $\mathcal{T}$ as
\begin{equation}
(f|g) = \frac{1}{T}\int_0^T f^*(t)g(t)dt,
\end{equation}
and the scalar product on $\mathcal{H}\otimes \mathcal{T}$ as
\begin{equation}
\langle  \langle \cdot |\cdot \rangle \rangle = \frac{1}{T}\int_0^T \langle \cdot |\cdot \rangle dt.
\end{equation}
This definition coincides with the usual definition of the scalar product on a tensor-product space. Indeed, for two factorized states $|\Psi_1\rangle \rangle = f_1(t) |\phi_1\rangle $ and $|\Psi_2\rangle \rangle =f_2(t) |\phi_2\rangle $, with $f_1, f_2 \in \mathcal{T}$ and $|\phi_1\rangle, |\phi_2\rangle$ time-independent, we have:
\begin{equation}
\langle \langle \Psi_1|\Psi_2 \rangle \rangle = \langle \phi_1|\phi_2\rangle \frac{1}{T}\int_0^T f_1^*(t)f_2(t) dt = \langle \phi_1|\phi_2\rangle (f_1|f_2).
\end{equation}
A natural basis on $\mathcal T$ is obviously $\phi_n(t) = e^{-in\omega_dt}$, for which we use the notation $|n)$. By analogy with usual Dirac notations, we will also write $\phi_n(t) = (t|n)$. Let $\{|\mu\rangle\}$ denote a basis of $\mathcal{H}$, the vectors  $|\mu,n\rangle\rangle = |\mu\rangle\otimes|n)$ then form a basis of $\mathcal{F}$ and the projection on this basis coincides with the Fourier transform. In other words, with these notations, any periodic state vector $|\psi(t)\rangle $ of $\mathcal{H}$ is represented in $\mathcal{F}$ by a vector $|\psi\rangle \rangle$ whose components are given by
\begin{equation}
\langle \langle \mu,n|\psi\rangle \rangle = \frac{1}{T}\int_{0}^Te^{in\omega_pt} \langle  \mu|\psi(t)\rangle = \langle \mu|\psi^{(n)}\rangle.
\end{equation}
where $|\psi^{(n)}\rangle$ is the $n^{\mathrm{th}}$ Fourier component. 

Coming back to the eigenvalue problem of Eq. (\ref{eqPeriodic}), it has a time-independent formulation in $\mathcal{F}$ and  can be written as
\begin{equation}\label{floquetEigFSpace}
\tilde{A}|p_{\alpha}\rangle \rangle = \epsilon_{\alpha}|p_{\alpha}\rangle \rangle,
\end{equation}
In the basis introduced above, the matrix elements of the operator $\tilde{A}$ are given by
\begin{equation}
\langle \langle \alpha,n|\tilde{A}|\beta, m \rangle \rangle = A^{(n-m)}_{\alpha \beta} - n\omega_d\delta_{nm}\delta_{\alpha \beta}.
\end{equation}
If $\tilde{A}$ is diagonalizable, we can find a basis of eigenvector in $\mathcal{F}$. Since $\mathcal{F}$ is infinite dimensional, let us label the eigenvalues and eigenvectors of Eq.~(\ref{floquetEigFSpace}) with a double index, $\{|p_{\alpha,k}\rangle\rangle, \epsilon_{\alpha,k}\}$, where $1\leq \alpha \leq N$ and $k\in \mathbb{Z}$. In principle, for every such eigenvector and eigenvalue, one can define a solution of Eq.~(\ref{eqDiffFlo}) given by 
\begin{equation}
|X_{\alpha,k}(t)\rangle = e^{\-i\epsilon_{\alpha,k}t}(t|p_{\alpha,k}\rangle\rangle.  
\end{equation}
However, we know from the theory of ordinary differential equations that only $N$ such functions are linearly independent. This is reflected in the following relation between eigenvalues and eigenvectors in $\mathcal{F}$ : let $p_{\alpha,0}$ denote the eigenfunctions whose eigenvalue satisfies $|\epsilon_{\alpha,0}| < \omega_d/2$, the other eigenvalues and eigenvectors are given by
\begin{align}
\epsilon_{\alpha,k} &= \epsilon_{\alpha,0}+ k\omega_d,\\
|p_{\alpha,k}\rangle \rangle  &=  \sum_{n= -\infty}^{\infty} |p_{\alpha,0}^{(k+n)}\rangle\otimes|n),
\end{align}
or equivalently,
\begin{equation}\label{relP}
(t|p_{\alpha,k}\rangle\rangle = e^{ik\omega_pt}(t|p_{\alpha,0}\rangle\rangle.
\end{equation}
This simply means that for any $k\in \mathbb{Z}$, $|X_{\alpha,k}(t)\rangle = |X_{\alpha,0}(t)\rangle$.

The advantage of introducing the Floquet space is that Eq.~(\ref{floquetEigFSpace}) is time-independent. The dynamics in $\mathcal{H}$ can therefore be computed in the following way:  let $|X_0\rangle\rangle$ denote a periodic function satisfying $(t|X_0\rangle\rangle|_{t= 0} = |X(0)\rangle$ (a possible choice is the constant function $|X(0)\rangle\otimes |0))$. The time evolution of $|X\rangle$ is then given by,
\begin{equation}\label{dynaFlo}
|X(t)\rangle = (t|e^{-it\tilde{A}}|X_0\rangle \rangle.
\end{equation}
The freedom in the choice of $|X_0\rangle\rangle$ comes from the infinite dimension of $\mathcal{F}$. Let us prove that it has no consequence on the dynamics in $\mathcal{H}$. For any initial vector $|X_0\rangle\rangle$ we can introduce the following decomposition
\begin{equation}
|X_0\rangle \rangle = \sum_{\alpha,k} \lambda_{\alpha,k}|p_{\alpha,k}\rangle \rangle.
\end{equation}
The initial condition then reads,
\begin{equation}
|X(0) \rangle = \sum_{\alpha,k} \lambda_{\alpha,n}(t|p_{\alpha,n}\rangle \rangle|_{t=0}.
\end{equation}
Using Eq.~(\ref{relP}), we find
\begin{equation}
|X(0)\rangle = \sum_{\alpha= 1}^N \lambda_{\alpha}(t|p_{\alpha,0}\rangle\rangle_{t=0}
\end{equation}
with $ \lambda_{\alpha}= \sum_{n = -\infty}^{\infty} \lambda_{\alpha,n}$. This last decomposition is unique since the functions $e^{-i\epsilon_{\alpha,0}t}(t|p_{\alpha,0}\rangle\rangle$ form a basis of solutions of Eq.~(\ref{eqDiffFlo}). Therefore, the coefficients $\lambda_{\alpha}$ do not depend on the choice of $|X_0\rangle\rangle$. Moreover, they completely determine the dynamics. Indeed, using again Eq.(\ref{relP}) we can write
\begin{equation}\label{dyna}
|X(t)\rangle = \sum_{\alpha = 1}^N\lambda_{\alpha}e^{-i\epsilon_{\alpha,0}t}(t|p_{\alpha,0}\rangle\rangle.
\end{equation}
Similarly, Eq.(\ref{dynaFlo}) can be extended to any initial time $t'$,
\begin{equation}\label{evolFloGen}
|X(t)\rangle = (t|e^{-i(t-t')\tilde{A}}|X(t'),0\rangle \rangle,
\end{equation}
where we have use the notation  $|X(t'),0\rangle \rangle = |X(t')\rangle\otimes|0)$.  Equation (\ref{evolFloGen}) thus defines the propagator $U(t,t')$ such that $|X(t)\rangle = U(t,t')|X(t')\rangle$. The matrix elements of $U(t,t')$ in the basis $\{|\mu\rangle\}$ then read,
\begin{equation}
U_{\mu,\nu}(t,t') = \sum_{\alpha,n,m} \langle \langle \mu,m |p_{\alpha,n}\rangle\rangle \langle \langle p_{\alpha,n}|\nu,0\rangle \rangle e^{-i\epsilon_{\alpha,n}(t-t')-im\omega_dt}.
\end{equation}
%
%
\section{Spectral properties of the Floquet-Liouville operator}
\label{app:meta}
In this appendix we prove the following properties of the periodic functions $R_{\alpha,k}(t)$ and $L_{\alpha,k}(t)$ introduced in the main text as left and right eigenfunctions of the operator $\mathscr{L}(t)-\partial_t$:
\begin{enumerate}
\item
if $\Omega_{\alpha,k}$ is an eigenvalue such that $\mathrm{Re}[\Omega_{\alpha,k}]\neq 0$, then $\mathrm{Tr}[R_{\alpha,k}(t)] = 0$ for all $t$.
\item
if $\Omega_{\alpha,k}$ in an eigenvalue, $\Omega_{\alpha,k}^*$ is also an eigenvalue and the corresponding eigenfunction is  $R_{\alpha,k}^{\dagger}(t)$.
\item
if $\Omega_{\alpha,k}$ is a real eigenvalue, $R_{\alpha,k}(t)$ and $L_{\alpha,k}(t)$ can both be chosen Hermitian.
\end{enumerate}
We assume that the operator $\mathscr{L}(t)$ is of Lindblad form, i.e. $\mathscr{L}(t)\rho = i[H(t),\rho] +1/2\sum_j (2C_i\rho C_i^{\dagger} - \rho C_iC^{\dagger}_i- C_iC^{\dagger}_i\rho) $, for some jump operators $C_i$.

\textit{Proof of 1.} This property follows from the fact that $\mathscr{L}$ is trace preserving: for any time $t$ and any matrix $\rho$, we have $\mathrm{Tr}[\mathscr{L}(t)\rho] = 0$. Injecting this relation into the eigenvalue equation satisfied by $R_{\alpha,k}(t)$ we find
\begin{equation}\label{traceEq}
\partial_t \mathrm{Tr}[R_{\alpha,k}(t)] = -\Omega_{\alpha,k} \mathrm{Tr}[R_{\alpha,k}(t)].
\end{equation}
In addition, $\mathrm{Tr}[R_{\alpha,k}(t)]$ must be periodic, (just as $R_{\alpha,k}(t)$). As a result, if $\mathrm{Re}[\Omega_{\alpha,k}]\neq 0$, the only solution to Eq.~(\ref{traceEq}) satisfying this condition is $\mathrm{Tr}[R_{\alpha,k}(t)] = 0$.

\textit{Proof of 2.} Due to the Linblad structure, the operator $\mathscr{L}(t)$ is invariant under Hermitian conjugation. More precisely, for any matrix $\rho$ we have
\begin{equation}
(\mathscr{L}(t)\rho)^{\dagger} = \mathscr{L}(t)\rho^{\dagger}.
\end{equation}
The result then follows by taking the Hermitian conjugate of the equation obeyed by $R_{\alpha,k}(t)$. We directly find
\begin{equation}
(\mathscr{L}(t)-\partial_t)R^{\dagger}_{\alpha}(t) = \Omega^*_{\alpha}R^{\dagger}_{\alpha}(t).
\end{equation}

\textit{Proof of 3.} Let $\Omega_{\alpha,k}$ be a real eigenvalue and $R_{\alpha,k}$ a corresponding eigenfunction. We deduce from Prop. 2. that $R^{\dagger}_{\alpha,k}(t)$ is also an eigenfunction with the same eigenvalue. Hence, if $R'_{\alpha,k} = 1/2(R_{\alpha,k}(t)+R^{\dagger}_{\alpha,k}(t))$ is not constant and equal to zero, then it is a proper Hermitian eigenfunction. In the case were $R'_{\alpha,k}(t) = 0$, then $iR_{\alpha,k}(t)$ is an Hermitian eigenfunction.
Suppose now that $R_{\alpha,k}(t)$ is Hermitian. In terms of Fourier components, this is equivalent to $R_{\alpha,k}^{(-n)} = R^{(n)\dagger}_{\alpha,k}$. Let us show that the corresponding left eigenfunction $L_{\alpha,k}(t)$ is also Hermitian. Given its expression in Floquet space, $L_{\alpha,k}(t)$ is uniquely defined by the following set of relations involving its Fourier components,
\begin{align}
 \sum_n\mathrm{Tr}[L^{(n)\dagger}_{\alpha,k}R^{(n)}_{\beta,l}]  &= 0 \quad \mathrm{for }\quad \beta\neq\alpha, l\neq k ,  \\
 \sum_n\mathrm{Tr}[L^{(n)\dagger}_{\alpha,k}R^{(n)}_{\alpha,k}]  &=  1.
\end{align}
From the fact that for every $\beta$ and $l$, $R^{\dagger}_{\beta,l}$ is also an eigenfunction, we find that
\begin{align}
 \sum_n\mathrm{Tr}[L^{(n)\dagger}_{\alpha,k}R^{(-n)\dagger}_{\beta,l}]  &= \sum_n\mathrm{Tr}[L^{(-n)}_{\alpha,k}R^{(n)}_{\beta,l}] &=  0.
\end{align}
Similarly, using the relation $R_{\alpha,k}^{(-n)} = R^{(n)\dagger}_{\alpha,k}$, we have
\begin{align}
 \sum_n\mathrm{Tr}[L^{(n)\dagger}_{\alpha,k}R^{(-n\dagger)}_{\alpha,k}] &=  \sum_n\mathrm{Tr}[L^{(-n)}_{\alpha,k}R^{(n)}_{\alpha,k}]  &=  1.
\end{align}
Combining these last two results, we see that the function $L(t)^{\dagger}_{\alpha,k}$, defined in terms of Fourier components by $(L^{\dagger}_{\alpha,k})^{(n)} = L^{(-n)\dagger}_{\alpha,k}$, satisfies the same set of relation as $L_{\alpha,k}(t)$.  Hence  $L_{\alpha,k}(t) = L^{\dagger}_{\alpha,k}(t)$.

\end{document}